# Appraisal of Social Learning Potentials in Some Trending Mobile Computing Applications

Abubakar Sadiq BAPPAH
*School of Technology Education
Abubakar Tafawa Balewa University,
P. M. B., 0248, Bauchi, Bauchi State Nigeria*
`Tel: (+234)8030462741`

*Abstract-* New Mobile technologies have created a new social dimension where individuals can develop increased levels of their social awareness by keeping in touch with old friends, making new friends, dispense new data or product, and getting information in many more aspects of everyday lives, making one to become more knowledgeable which is very beneficial especially for students. Social networks, in particular enable users to share and discuss common interests and provide infrastructures for integrating various user experiences: synchronous and asynchronous communication, game-playing, sharing links and files. The trend of using social networks and social media to deliver and exchange knowledge could bring a new era of social learning in which learners make use all four language skills of reading, writing, listening and speaking. Unlike a traditional e-leaning paradigm with pre-defined curriculum and standard textbooks, social knowledge could be aggregated on demand, just in time, and in context of engaging challenges from social networks, making learning more exciting, social and, game-like experience. Social learning environment engages the learners in discussion, collaboration, exploration, production, discovery and creation. Schools should harness this to develop group collaboration skills and even project based learning activities that span subjects and grade levels. This paper focuses on technologies and avenues of using mobile phones in supporting open social learning among undergraduate students in Nigeria.

## I    INTRODUCTION

As technology and market evolves, these devices become more pervasive and integrate larger parts of our everyday life. Nevertheless, it remains that the integration of technology in our lives is not the result of a secret process but of a complex social one. As we are moving toward a ubiquitous environment, a central question is whether and how mobile devices can be used to support social learning processes in the near future. Students grew up in the technology era and social networking is now just a part of their daily routine as they exchange ideas, feelings, personal information, pictures and videos at a truly astonishing rate. Social networks, such as Facebook and Twitter capture vast amounts of implicit knowledge, common beliefs, preferences, and experiences that could be potentially empower users to learn from each other and together. This implies social learning, that is, settings where students can learn from each other and interact with several groups. Because of the fast development of information technology in education, many high quality online courses and web-based educational programs have been adopted by both formal and non-formal education [1]. Also, with the support of information technology, the formal education gradually becomes more and more flexible in the mode of delivery than previously, not limited by the mode of face-to-face instruction. It seems that the distinction between formal and non-formal education inevitably becomes unclear. In an e-learning context, social computing is about students becoming the creators as well as the consumers of content. In a formal setting, students can be encouraged to use social computing technologies to share their experiences and collaborate on assignments and projects. In informal situations, people will be able to find great treasuries of information on almost any imaginable topic and contribute their own knowledge to it. Using mobile devices for learning is the logical next step for e-learning [2].

In general, the Internet and social networking sites provide an outlet for students to express themselves in their own unique ways. In addition, schools have commenced the use of these sites to promote education, keep students up to date with assignments, and offer help to those in need. This is due to the fact that the social networks can be used for various academic activities such as communicating with the faculty and university authority, communicating with lecturers and supervisors, making academic discussions with classmates and chatting with friends in respect to topics of educational interest. Students use these sites as tools to obtain information and resources for graduation preparation, research purposes, scholarship search, advance training, online schooling program, tips on exercise and nutritional requirements online library and e-books access and for future planning. However, there are always two sides to every story. The risks and dangers of Internet usage are constantly flooding television shows, newscasts, and magazines, always warning parents to educate parents on teen Internet behaviors [3]. Social networking sites, as well as other new forms of communication technology, are also a concern to many school professionals/lecturers because of the level of distraction they create within the school and the growing tendencies of sharing inappropriate information or disclosing too much information.

## II   EVOLUTION OF SOCIAL NETWORKS

In the early l990's online communication technologies were introduced to the public in forms such as e-mail and chatrooms [4]. America Online (AOL) Instant Messaging (shortened to AIM) was one of the first online communication technologies that encouraged users to communicate with existing friends in real-time. In order to have an online conversation with another person, a user would have to send an add invitation to another user, which would then have to be accepted by that person, acknowledging that he or she was an existing friend. Each user creates his or her own screen name which, for most, was either the user's first name followed by numbers or a fictional name or word. When using AIM, each user has his or her own buddy list which displays the current screen names of the people he or she knows and is able to chat with. In the early stages of AIMs popularity, users would have to use a phone line in order to use the program, as this was the only means of gaining Internet access. Once the Internet was able to be accessed through cable and broadband connections, users would be able to keep the Internet running as long as they wanted, which allowed them to send and receive messages throughout the day. This acted as a way to communicate with others even when a friend or family member was not currently sitting at their computer. Away Messages were created to inform others of what a user is doing when not currently online. If a user tried to send an instant message to a friend, these away messages would be sent back to that user, alerting him or her that the message was received, but that user is away from the computer at the moment.

## III   MOBILE DIGITAL SOCIALIZATION

The growing trend of online social networks proliferation in recent years has attracted hundreds of millions of Internet users participate in social networks, form communities, produce and consume media content in revolutionary ways. Students have especially embraced this new way of communicating for entertainment and also academic purposes. Though only a few have gained worldwide publicity and attention, Social Networking sites have attracted and fascinated tens of millions of Internet users and it is estimated that there are over 200 different sites that are used for social networking [5]. Accordingly, females are more likely than males to engage in social networking and males tend to browse for females. Social Networking is one of the primary reasons that many people have become avid Internet users; people who until the emergence of social networks could not find interests in the web. The idea behind most of this phenomenon, as with many websites, is to help people feel socially connected and part of a community, even though they may be sitting home alone at their computer [6]. Social networking websites function like an online community of internet users. Depending on the website in question, once a community member is granted access to a social networking website he begins to socialize. This socialization may include connecting with other people they know through school, work, or an organization, or they may meet complete strangers from all over the world. They do this by searching for people and adding them as friends so that they may share information with them and other networks that those people may be a part of. Though there are several options for privacy on these sites, research has shown that the public aspect of sharing information is what draws many to join and participate [5]. Social Networking avenues have influenced the way we deal with others, entertain and actually live. Commonly, digital natives and digital immigrants as well now have devices that send and receive email on the go. Such devices are also used, in most cases, for voice communication, data communication, instant messaging amongst others. A click of a button may mean the loss or gain of a friendship, and a friendship on a social network may be with someone who is not a friend in real life.

## IV   MOBILE SOCIAL LEARNING

In recent years there has been a groundswell of interest in the use of mobile devices as a consequence of two intertwined and digitally converging technologies of mobile computing and Web 2.0 advancements. Specifically, the rate of adoption of mobile technologies in Africa's developing countries is amongst the highest in the world and there may be almost 300Million mobile users in Africa [7]**.** The growing trend of online social networks proliferation in recent years has attracted hundreds of millions of Internet users participate in social networks, form communities, produce and consume media content in revolutionary ways. Mobile social learning takes place when students are supported in their learning activities with portable computational devices. Mobile technologies have increased the opportunities for informal learning. Technologies such as mobile phones and Personal Digital Assistant (PDA) offer students with opportunities to communicate and access information at any time and from any location. In addition, these tools allow increased autonomy and control meaning that students can choose where, when and how they learn. From an economic perspective, mobile learning reduces costs of infrastructure and physical instructional materials, since it does require a cyberspace only. It has been highly recognized as a strategic tool that has the potential to enable global access to educational materials and prepare them for change [8]. This certainly challenges the traditional classroom with a fixed location, a teacher, a curriculum and assessment. Literature has shown that mobile social networks provide a casual place of learning, encourage students to express their own thoughts, provide effective collaboration and social learning experiences [9]. Social learning potentials of applications on mobile devices can be explored dynamically in different contexts of situated learning activities:

1. Communication and Microblogging: These are general applications which can support communication and collaboration among students and others [10]. Relevant social networking applications range from common e-mailing, SMS and voice calling platforms to special online as well as offline packages messengers. Mobile device users could perform collaborative activities which include sharing of files, data, and participation in forums, wikis and blogs. For example, Thornton and Houser [11] studied students' patterns of mobile phone use and found that e-mail was the most frequently used application, with 66% using it to exchange information about course content. Other areas of useful support include the use of SMS facility, instant messengers, voice calls, and data calls where students can discuss course contents.

2. Representational and Social graphs: Representational would enable students to create representations to illustrate their ideas using concept maps. These maps were shared among classmates and used as a means to promote academic discussion. A graph is a mathematical abstraction for modeling relationships between things. A graph is constructed from nodes (the things) and edges (the relationships). This mathematical tool that can model natural and artificial systems such as economy, deceases, power grids, etc. has been used by the anthropologists, sociologists and other humanities oriented academics. Similarly, graph analysis on social platforms can enhance collaboration and teamwork among social learners.

3. Library and e-Resources: These are facilities that can access to information from vast sources such web based dictionaries, search engines, virtual libraries, professional networks, media exchange facilities, and academic links. Although, these activities are more often performed individually they can also involve sharing of media and text based resources with others. This form of social learning reflects both a behaviourist and constructivist approaches since mobile users are to explore and construct their own understanding of issues. Brandt, Hillgren and Bjorgvinsson [12], for example, found that medical students used their mobile phones to access needed information while on training and were able to review content and notes.

4. Mobility and Geolocation: Mobility is the ability to access information and collaborate with others from anywhere and at anytime. Convergence is making this possible, with music players, wi-fi connectivity, video cameras, GPS units, and live television capable of running on a single device, often a mobile phone. The days of carrying a separate phone, camera and music player are over. Indeed, use of the word *phone* is fast becoming obsolete, preferring to refer to these new gadgets as mobile communication devices, or digital assistants. Location-based social networks allow members to share their location through GPS, Bluetooth, email or text messaging. The member of the network may also add comments about restaurants, allow friends to know where you are going, share information, or find friends that are few blocks away or even in the Café across the road. Several mobile-only social networks have emerged, all with unique features that would potentially attract users. However, there is one feature that every mobile social network should have and this is physical presence detection and information exchange.

5. Social Business Solutions: Another area with huge growth potential is the use of social networks and/or social networking techniques for business whether it is for improving communications, for marketing or by deriving business intelligence. There are now many different solutions in the market for "Social Business Solutions" and large IT companies and consulting forms are starting to create new departments in this area [7]. Businesses are slowly coming to the realization that traditional communication methods are often less effective than social networking. These sites offer services and opinion sharing by the users. One of the open issues is to establish methodologies and tools to distinguish between real and fake opinions on those social sites, which is applicable to social networks in general.

6. Multimedia and Capture Tools: These are support data gathering and information that may provide further learning opportunities. Students may walk around the city and take notes or record information using their mobile device cameras. On visits to museums, industries and other study sites students can use their mobile phones to take photos, or record observations. These can be shared later with classmates and family. However, recording one's voice and making video clips are fast gaining acceptance among social network users.

7. Social Television: Social Television stands for technology that provides social interaction in the context of watching TV-programs or related to television content. It is a very active area of research and was named one of the 10 most important technologies by the MIT [7]. Social Television is a fast growing market sequel to the rise of social networks that already encourage constant connection between members of the network and the creation of likely minded groups. Social television is connecting viewers with their friends, families and people with same interests giving them a space to discuss and exchange recommendations. Today's digital natives access the internet more oft en than they watch television and a growing number among them is interested in having more social features integrated into their TV-experience.

8. Social Gaming: Social Gaming is a term for games that are based on social interaction. By augmenting the game logic with social aspects players have to deal with each other in various ways to advance throughout the game. Recent technological advancement in mobile communication and computing technologies created the ground for a new field of games with social interaction as the main focus. Shaffer [13] describes a variety of educational learning experiences that virtual reality could present to middle school, high school and even college students. For instance, students will choose whether they will play as red blood cells, white blood cells, viruses, or anti-viral drugs to learn how viruses affect the body, and how to stop them. A major hindrance to exploitation of social network data in this regards is the fragmentation of the population of social network users into numerous proprietary and closed social networks. This issue is compounded by the fact that each new game or media application tends to build its own social network around it rather than building upon the rich data available about existing social relationships. Also applications are oft en restricted to execute within the confines of specific social network platform. A major research challenge, therefore, that would benefit the exploitation of social network graphs for future media networking, is in finding solutions to open up social network platforms to allow cross-platform information exchange and usage.

9. Assessment and Project Management Tools – Assessment tools refer to the uses of mobile phones or PDA for students to answer examination questions, tests, or quizzes [9]. Instructors can post a series of multiple choice questions to a portal which students could access via hand held devices. After completing the questions, students would receive a score and could review their answers. With regards to analytical capabilities, it facilitates manipulation of data using trending graphic calculators or other analytical tools. As for the task management tools, they provide time management support in forms of timers, calendars, tasks managers, etc. All these administrative tools could support informal learning efforts.

V CONCLUSION

Taking the trends discussed in this paper into account it can clearly be seen that the education system should change to adapt to modern requirements and to incorporate new technologies. By incorporating these technological trends into the educational system a higher quality education can be provided at a cheaper cost and spread over a larger segment of the population. However, the initial cost of incorporating these technologies into any school, yet alone a school in rural Africa, will be extremely high. While social networking sites are valid resources for students, educators and administrators, it must be acknowledged that social networking, as is the case with most technology, comes as a mixed blessing. More emphasis should be placed on how educators and parents can help guide students to enjoy the benefits of social networking while also recognizing the problems they may create. The level of students' participation in social networking sites is on a daily rise, when properly harnessed those applications would foster mutual benefits among the students and their teachers as well.